# Incorporating an economic approach to production in a health system model

Martin Chalkley[1*], Tim Colbourn[2], Timothy B. Hallett[3], Tara D. Mangal[1,3], Margherita Molaro[3], Sakshi Mohan[1], Bingling She[3], Paul Revill[1] and Wiktoria Tafesse[1]

* Lead author and corresponding author
All other authors are listed in alphabetical order
[1] Centre for Health Economics, University of York, York, UK
[2] MRC Centre for Global Infectious Disease Analysis, Jameel Institute, School of Public Health, Imperial College London, London, UK
[3] Institute for Global Health, University College London, London, UK

**Corresponding author details:**
Name: Martin Chalkey
Email: martin.chalkley@york.ac.uk
Phone: +44 (0)1904 321433

**Funding Sources:** The Thanzi la Mawa project is funded by Wellcome (223120/Z/21/Z). TBH, TM, MM and BS acknowledge funding from the MRC Centre for Global Infectious Disease Analysis (reference MR/R015600/1) along with funding through Community Jameel.

## Abstract

As computational capacity increases, it becomes possible to model health systems in greater detail. Multi-disease health system models (HSMs) represent a new development, building on individual level epidemiological models of multiple diseases and capturing how healthcare delivery systems respond to population health needs. The Thanzi la Onse (TLO) model of Malawi is the first of its kind in these respects. In this article, we discuss how we have been bringing economic concepts into the TLO model, and how we are continuing to develop this line of research. This has involved incorporating more sophisticated approaches to account for the effects of the unavailability of healthcare workers, and we are working towards establishing the role of different forms of ownership of healthcare facilities and different management practices. Not only does this broad approach make the model more flexible as a tool for understanding the impact of resource constraints, it opens up the possibility of analysing considerably richer policy scenarios; for example establishing an estimate of the health gain that could be achieved through expanding the workforce or reducing healthcare worker absence.

**Keywords:** Health System Models, Production functions, Production theory, Human resources for health, Health System Strengthening, Conceptual Framework

# Introduction

Significant advances have been made by extending disease-specific models to encompass all diseases. By applying dynamic mathematical and computational methods, Multi-disease health systems models (HSMs) quantitatively explore the underlying dynamic and feedback relationships between diseases, access to preventative and curative treatment and population health[1] . HSMs are a valuable tool for health systems researchers and decision-makers alike, as they can describe, predict, and quantitatively capture population health and be used to forecast the demand for healthcare due to the spread of diseases and population changes. The use of HSM in the economic evaluation of healthcare system interventions can support cost-effective investments[1,2].

Given the evolution of HSM from a focus modelling the progression of specific diseases, it is not surprising that models are typically strong on the determinants of ill health, but relatively weak in respect of the ways in which healthcare is delivered and thereby impacts on health. As reviewed by Chang et al.[1] there is often a black-box approach to the role of healthcare systems in HSM, and no clear articulation of the role of different inputs (healthcare professionals, medicines and facilities), how they are defined and measured, possible dynamic and interactive relationships between them or the outcomes and outputs (intended or otherwise) they produce. Examples of this black box approach are Atun et al.[3–5], Kok et al.[6] and Dangerfield et al. [7] Some operational components of health systems are captured in the approach of Langley et al.[8] and Lin et al.[9] but these only model a single disease. Against this background we describe how we are developing, both conceptually and practically, an economics-based framework to account for the role of healthcare in a genuinely multi-disease model – the TLO whole system model for Malawi[10].  Our goal is to enhance the  relevance, policy usefulness and applicability of the TLO model.

Our conceptual viewpoint is the economics of healthcare production that provides a framework for how to model the building blocks of healthcare provision. It distinguishes between individuals with a perceived healthcare need who make choices of whether and where to seek care (demand) and healthcare providers (which can be individual doctors, or 'organisations', individual clinics or hospitals or broader corporate bodies) who determine what sorts of care to offer, who to prioritise for treatment, and what sorts of treatment to give (supply).  This provides a basis for incorporating practical constraints on which individuals with which diseases will receive treatment. This is in stark contrast with a single disease setting, which cannot account for the overall volume of demand placed on the healthcare system, and therefore cannot enforce limitations on overall healthcare delivery.

Just how important  might such factors be for improving the health of the population?  As with many other questions a model can provide a guide.  As we have developed the production side of the TLO model we have been able to simulate the effects of changing production parameters. Some preliminary estimates and calculations are summarised in Figure 1 below indicating that between 10% and 20% more overall health (measured in terms of DALYs averted) might be available through relaxing some of the existing resource constraints in the system. That in turn suggests that very substantial health gains may be possible through ensuring better functioning of the healthcare production system.

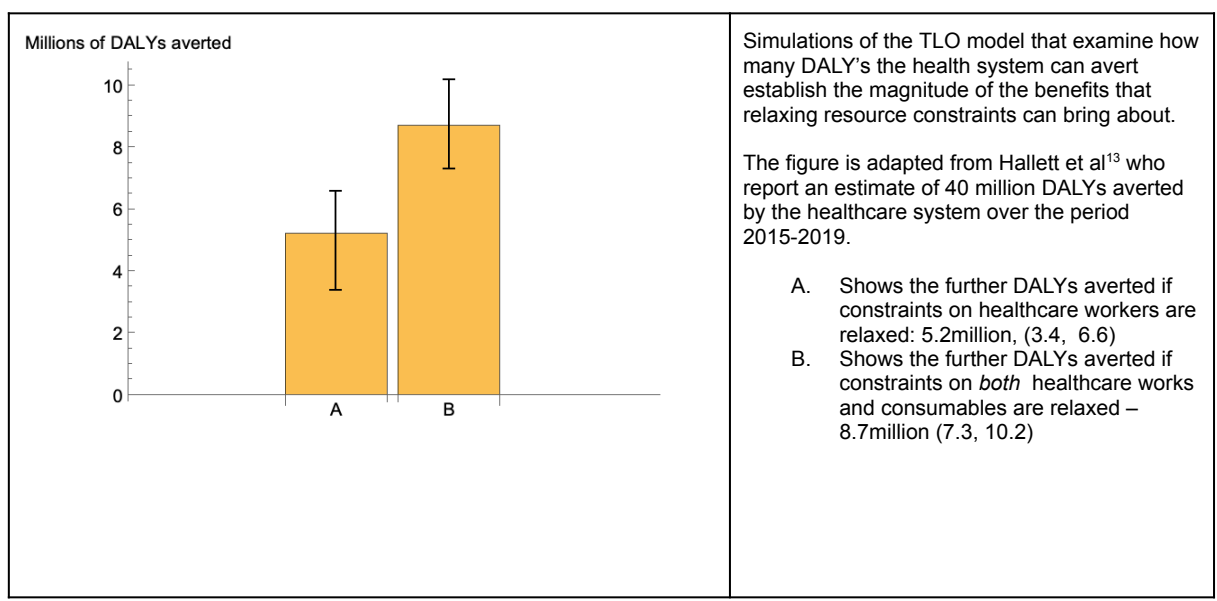

**Figure 1: The potential impact of relaxing healthcare resource constraints on population health**

In the remainder of this article we set out the foundational principles of the TLO whole system model of Malawi; summarise our progress in implementing an economic approach to healthcare production in the TLO Model so far, set out our aspirations going forward and discuss the value of our approach.

There are a number of areas of literature which influence our approach and which we contribute to. For example, the developments we describe can be seen as implementing the suggestions that future HSM research should take into account the dynamic connections between the interdependent components of health systems, such as financing, delivery, individual behaviour and community decision-making, while considering feedbacks, adaptations, and synergies [2]. Our economic focus also relates to a growing literature which proposes and describes the use of whole-HSM as a basis to assess the economic evaluation of health interventions [2,11]. Notwithstanding this, there has only been very slow advancement of the use of economics to inform HSM. Among papers identified in a review of health systems modelling [12] none explicitly incorporate economic concepts, theory, or parameters from the economic literature in their modelling framework and hence we start with a relatively blank canvas.

## The TLO model

The TLO model [11,13–15] is a whole country individual-based simulation engine used to track a population and health events in Malawi, through a number of modules. There are three main types of modules; core modules characterising basic processes (demography, lifestyle, contraception and health seeking behaviour), multiple disease modules describing disease dynamics, progression and effect of receiving or not receiving treatment (Tuberculosis, Cancer, etc.), and the health system module representing how health service capabilities are distributed across the population needs (generated by the disease modules).

The health system module represents the provision of public healthcare consisting of public facilities and publicly supported non-profit facilities managed by the Christian Health Association of Malawi (CHAM). This module represents healthcare provision capacity in terms of healthcare worker time, the availability of consumables and equipment, and bed capacity, specified at the level of each simulated representative facility and for each simulated day. The demand side is in turn primarily determined by the symptoms that an individual exhibits and their health seeking behaviour according to their income, education, characteristics and location.

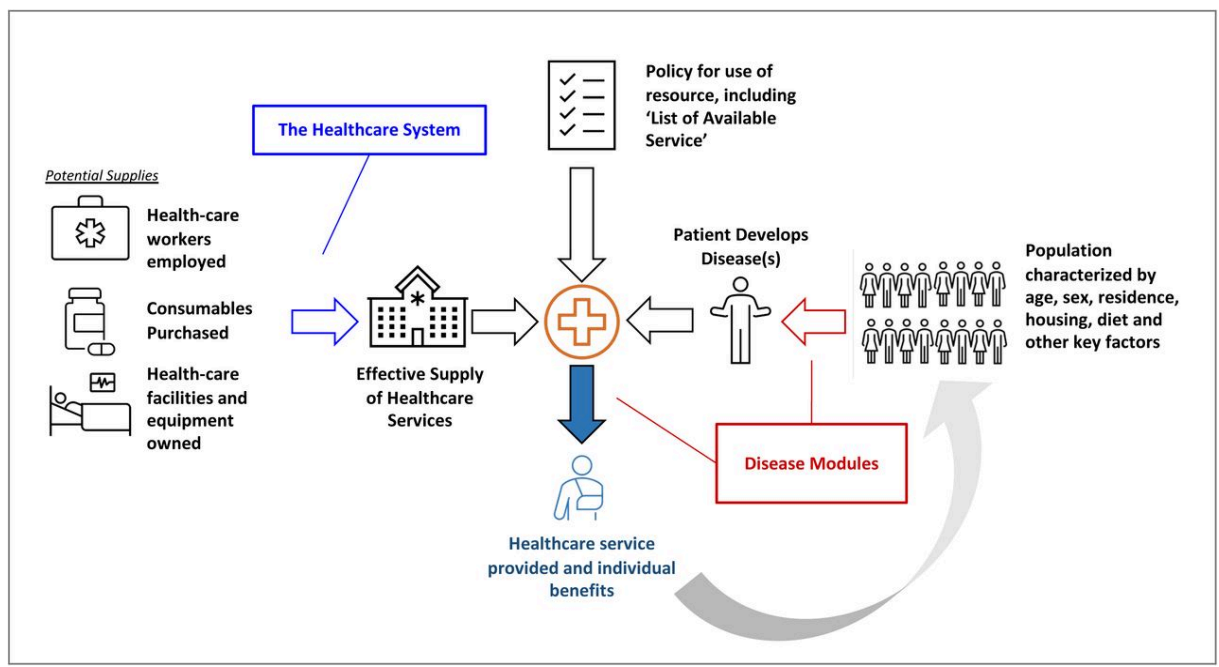

**FIgure 2: Schematic of the TLO Model** (source Hallett et al.[13])

Within the health system module, healthcare worker supply is defined as available patient facing time which is a product of the staff count and patient-facing time by health system level and healthcare worker cadre per district[16]. Individual facilities are not represented but rather grouped into sets of facilities of the same level of care and district. In terms of effective healthcare – taking into account whether the service delivery is appropriate – the TLO model includes disease specific diagnoses and treatment accuracy for each health system interaction event (HSI). Each HSI event is associated with a diagnosis or treatment given at the facility level through the interaction between demand and supply. The model represents the usage of diagnostics, including their imperfect performance, and effectiveness of treatments.

Our focus is on developing this model to account for health worker constraints, limitations on the supply of consumables and the effects of institutions and location in ways that are suggested by economic theory.

## Economics of healthcare production

A first key element of an economics-based approach is that the ability of any system to deliver output is the scarcity of all underlying resources. Hence in an economic approach the total production of a health care system will always be limited by the resources available to it, categorised as labour (the healthcare workforce), capital (physical healthcare facilities and medical equipment) and materials (drugs and other medical supplies).  It is usual to regard total production as being determined by the combination of these resources available and often that dependency is summarised in what is termed a production function[17].  Most formulations of production functions embody the idea of substitutability between inputs, so that the absence of one input can be compensated for by increasing the other inputs. How such possible combinations are managed depends on who is responsible for making decisions on production. Commonly used functional forms for production functions include the Constant Elasticity of Substitution, the Cobb-Douglas and the Leontief.[18]

In the view of economics production is about assembling and organising resources and that can be subject to a myriad of influences mediated by producing organisations - typically called firms. In healthcare we usually characterise production as being the remit of healthcare providers who serve as the analog of firms in that setting.

One influence on the institutions that we call providers, and hence a key determinant of production, is their intrinsic motivation and objectives. Providers can be for-profit, not-for-profit or government owned and each of these implies potentially different goals that will influence decision making. Providers can be viewed as determining how resources are deployed, and in the case of labour, how well (or otherwise) workers are motivated.

A second key element influencing the management of scarce resources in production – one that has been explored extensively in both high- and low-income country settings – is the role of extrinsic (financial) motivation and in particular how these might affect the impact of intrinsic motivation. Hence, there has been great interest in conditioning the payment of healthcare providers on certain aspects of performance, including the volume of treatments they engage in (activity-based finance) or the quality/effectiveness of the treatments they deliver (pay-for-performance).

In the light of the above, our continuing development of the health system module of the TLO model concentrates on better understanding and accounting for scarcity of healthcare resources, and better understanding and accounting for some of the determinants of how those resources are deployed, subject to the influence of providing organisations.

## Modeling Production in the TLO Model

### Incorporating production limitations

Starting from a position in which healthcare was assumed to be delivered regardless of whether the relevant healthcare worker was available at a facility, we have incorporated some key elements of the economics of production functions into the model.

A full specification of how healthcare could be delivered by varying each, or all of the required inputs; labour, capital (such as equipment and beds) and consumables.  However, a reasonable working assumption is that treatment requires both healthcare worker time and consumables, in given proportions that vary according to the disease being considered. This is the essence of what is termed a Leontief type production function[18,19] that has formed the basis of our approach. Using existing data on the mix of inputs required to produce treatment - very specifically the amount and composition of healthcare worker time - we have adapted the health system module to require that treatments can only be carried out if all the required inputs are available at a facility.

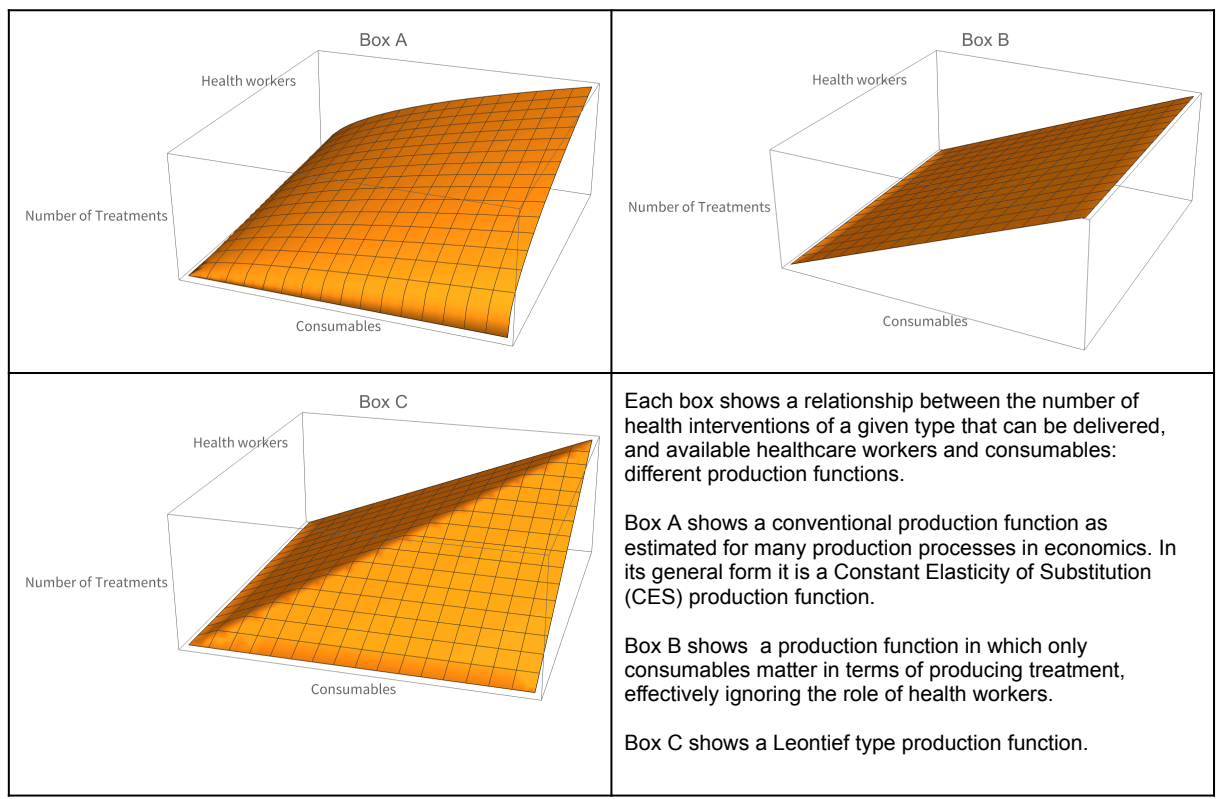

**Figure 3: Different production functions to characterise a health system module**

Box C in the figure shows an idealised (theoretical) Leontief production function. When implemented in the model the relationship between available inputs and treatments delivered is far more complex, because resource availability is constantly dynamically updated according to the requirements of a whole host of treatments that the health system is being asked to deliver. Graphically this results, as in Figure 4, in a surface perturbed in a highly non-linear way. Furthermore, in the model the surface has many dimensions, reflecting different consumables and different cadres of health workers.

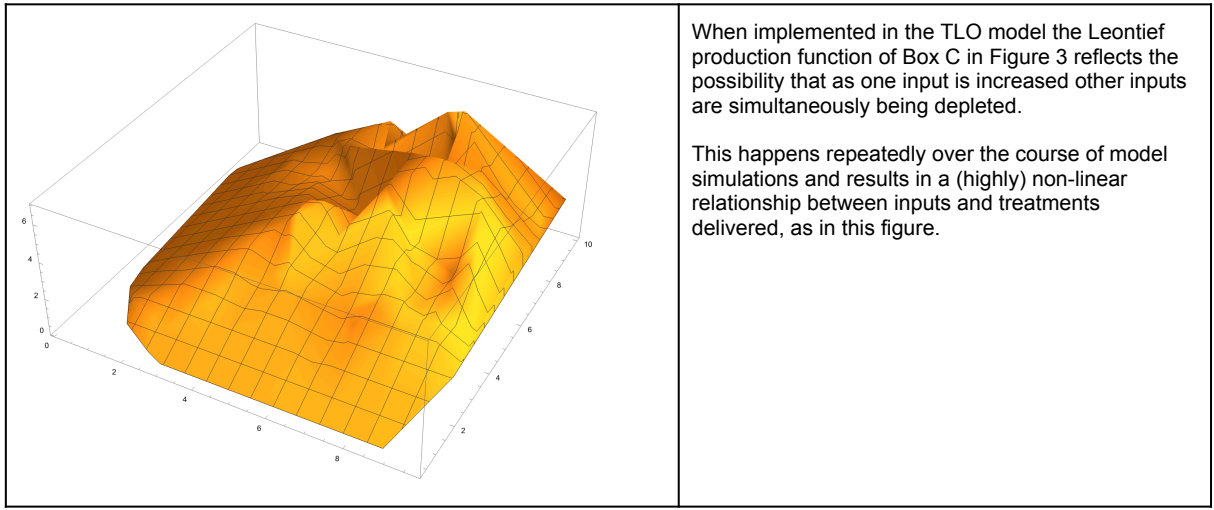

When implemented in the TLO model the Leontief production function of Box C in Figure 3 reflects the possibility that as one input is increased other inputs are simultaneously being depleted.

This happens repeatedly over the course of model simulations and results in a (highly) non-linear relationship between inputs and treatments delivered, as in this figure.

**Figure 4: The Leontief production function as implemented in the model which reflects dynamically changing resources**

In our first implementation of this production function approach we recognise the necessity of a healthcare worker being present, but do not impose a limitation on their actual use. This is equivalent to assuming that a worker can be utilised beyond their standard working hours (or equivalently that the treatment they deliver can somehow be compressed in time). We have developed this to specifically require that the healthcare worker has the necessary time available to treat a patient, having tracked the use of their time up to the point when that patient arrives.

The different assumptions underlying production are termed Modes. Mode 0 is the starting point of not requiring the relevant healthcare worker to be present. Mode 1 requires the presence of relevant health worker cadres, but does not account for any limit on their available time and Mode 2 captures a more plausible resource constraint in respect of healthcare worker time.

Comparison of Modes 1 and 2 in simulation studies provides an immediate indication of how limited healthcare worker time might be affecting the health of the population.

## Reflecting some of the influence of providers

### Ownership and organisational form

The modelled healthcare supply in the TLO model represents healthcare workers in public and non-profit health facilities managed by the government of Malawi and CHAM, respectively. Theoretical and empirical research, shows that healthcare ownership (generally categorised into four different types; public, private-for-profit, private-non-for profit and informal providers) may determine how much a provider cares about; profits, organisational mission, pro-poor service provision and public goods. Subsequently, these differences may impact which services are delivered[20,21] and the quantity and quality of health care provided.

We are presently engaged in empirical research to establish plausible parameter shifts that reflect differences between differently owned organisations.

The TLO model assumes that the healthcare delivery by healthcare workers is independent of the facility they work in. However, evidence shows that healthcare worker performance is not orthogonal to facility characteristics and that some portion of healthcare worker behaviour is facility specific[22,23]. For example, the size and scope of facilities seem to matter for healthcare delivery as increased facility size makes it more difficult to monitor worker effort[24,25]. An investigation of these influences is also work in progress.

Performance of the workforce

We have calibrated health worker productivity in the model to reflect the observed number of services delivered by staff. This is already an improvement over a rigid conception of a fixed time required to deliver services. What remains is to further generalise the model's conception of the determinants of health worker productivity. Again, we are using economic concepts of production and productivity.

Since organisations are the sum of the individuals that they have at their disposal, one avenue to consider is the role of the motivation of those individuals. There is a very specific focus on those responsible for delivering healthcare – the healthcare workforce – and what motivates them. Again it is traditional to distinguish between the goals and objectives of the individuals themselves, their intrinsic motivation, and the mechanisms that an organisation uses to try and align the decisions of individuals with the collective goals of the organisation. That includes a discussion of the pay, remuneration more generally, career progression prospects and so on.

The average staffing across facilities is far below the stated staffing norms for many countries in Sub-Saharan Africa (SSA) due to substantial absence of workers from health facilities[26]. Using data for multiple countries in SSA, Sheffel et al.[26] find that overall health worker absence was, on average, 34.7%, and similarly high across all types of health facilities, all cadres of health workers and in both urban and rural facilities. It has been shown that health worker absenteeism reduces the odds that patients, particularly young children, seek care in the public sector and receive malaria testing, and increases the odds of paying out-of-pocket for treatment[27]. We are engaged in research to establish the key economic factors inherent within the health system which influence the attendance of healthcare workers in Malawi

In respect of commitment of time, the TLO healthcare supply module assumes 1) that healthcare workers spend a certain amount of minutes per day seeing patients as per estimates from policy documents and expert consultations and 2) that a given healthcare interaction event (case) will require a certain amount of healthcare worker time based on estimates from expert consultations and Time & Motion Studies[16,28]. However, these estimates may be biased due to inaccurate reporting or the consequences of those studied being aware of their involvement – the Hawthorne effect[29]. Economics literature from LMICs shows that providers spend significantly less time seeing patients than modelled. They spend on average between 3-9 minutes per patient consultation[30–33] and in total between 40 minutes - 2 hours per day seeing patients. We have collected a large amount of time and motion data for a cross-section of all health workers and facility types in Malawi in 2024 that we are currently analysing to shed light on how much time is spent by health workers with and for patients with a full range of health conditions.

One way to encapsulate[21] these diverse influences on workforce performance is to recognise that the production functions shown in Figure 3, and assumed to be common across the healthcare system, are in fact specific to the setting in which healthcare is delivered. There may be shifts in these functions according to who owns a facility, where it is located, how it is managed and how large it is. This is illustrated in Figure 5.

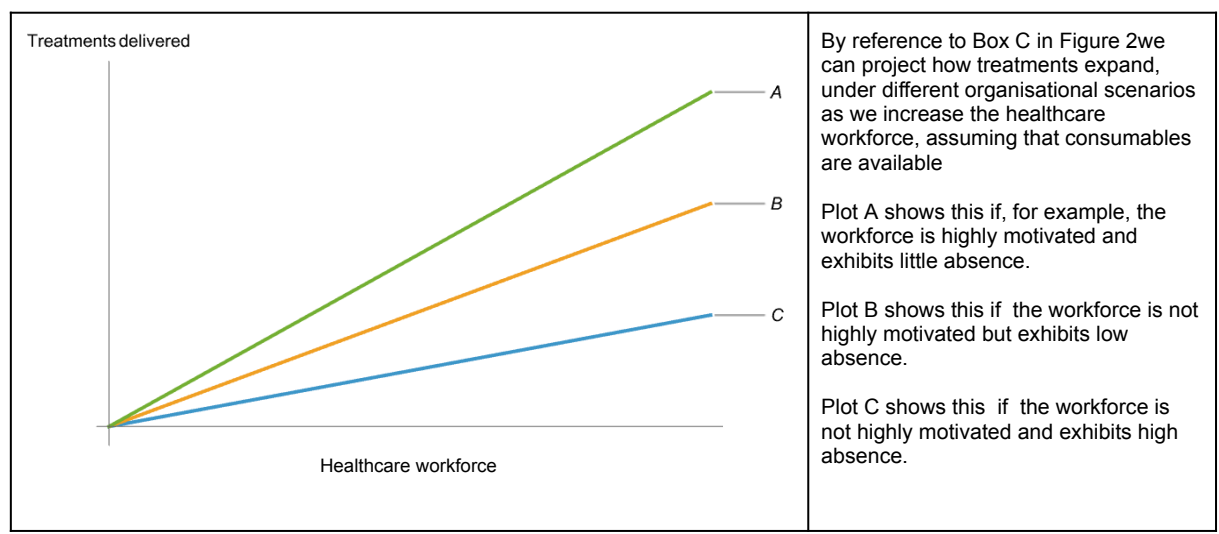

**Figure 5: The potential impact of motivation and absence on healthcare production**

As with Figure 3, the idealised linear representation of the production relationship in Figure 5 is translated into a nonlinear relationship in the model as other inputs are utilised.

The informational requirements to establish the existence and magnitude of the shifters in Figure 5 are huge. We are therefore progressing incrementally and picking low-hanging fruit where they appear. Examples are the role of ownership - where comparative data on differently owned facilities exist - and our investigation of variation in healthcare worker attendance, working time and quality of care across a broad suite of facility specific characteristics.

## Capturing the impact of consumables

In order to develop an economic model of production further it is important to have information on the availability of all inputs to the production function. Whereas, data on the staffing of health facilities provides a good basis for establishing the availability of healthcare workers, the availability of medical consumables is more problematic. By their very nature consumables become depleted in a random fashion according to use. How stocks are managed and replenished to deal with their usage is influenced by the policies and financial constraints of healthcare facilities.

The extent to which there is systematic variation in consumable availability across different facilities according to their size, ownership and management has been the subject of recent research[34] which has established important differences across different healthcare facilities in Malawi. This adds a further dimension by which the model of production can be refined, and also highlights a hitherto unexplored role for policy interventions to increase healthcare production; by understanding better the causes of stock-outs and where they are most likely to occur it should be possible to intervene in a targeted manner to reduce their incidence.

## Discussion

HSM models are developing into a valuable resource to assist decision makers in low-income countries who face difficult choices regarding which diseases to target, which healthcare services to fund, which health system building blocks to invest in and how to create conditions for the efficient use of scarce resources to achieve health sector objectives. Whilst no model is 'realistic', because by definition it is a highly simplified abstraction, a model that captures as much of the real-world constraints of a healthcare system is likely to both present a closer representation of reality and allow the evaluation of a wider range of alternative policy choices. Hence, our goal in incorporating the economics of production into the healthcare system module of the TLO model for Malawi is to increase its usefulness and relevance.

Whilst disease modelling has made huge strides forward in recent decades, the modelling of both healthcare seeking behaviour and healthcare production tends to be highly simplified. Economics has something to contribute in both of these domains but in this article we have focused on healthcare production because there is more readily available data to inform the development of the TLO model in this direction. A key avenue to explore in the future is the interaction between the availability of healthcare – supply – and healthcare seeking – demand[35].

The ideas and work we have reported here are facilitating the posing of questions and analysis that were previously not possible. We have focused on ensuring that the *structure* of the health system module in the TLO model is capable of incorporating emerging research findings by building a conceptual framework of production that accords with some core economic theory and concepts. That framework continues to be calibrated and informed by research findings concerning: the productivity of healthcare resources in Malawi, how productivity varies according to healthcare setting, facility ownership and facility management, healthcare worker availability and productivity and the availability of medical consumables.